\documentclass[amsmath,amssymb,aps]{revtex4-2}

\usepackage{graphicx}
\usepackage{hyperref}
\hypersetup{hidelinks}
\usepackage{bm}
\usepackage{dcolumn}
\usepackage{booktabs}
\usepackage{array}
\usepackage[dvipsnames]{xcolor}
\newcommand{\secondrev}[1]{\textcolor{blue}{#1}}

\newcommand{\Ep}{E_{\mathrm p}}
\newcommand{\lp}{\ell_{\mathrm p}}
\newcommand{\ii}{\mathrm{i}}
\newcommand{\dd}{\mathrm{d}}

\providecommand{\orcidlink}[1]{\href{https://orcid.org/#1}{\raisebox{0.2ex}{\tiny ORCID}~#1}}

\begin{document}

\title{Klein--Gordon oscillator with linear--fractional deformed mass-shell invariants in doubly special relativity}

\author{Abdelmalek Boumali}
\email{boumali.abdelmalek@gmail.com}
\thanks{\orcidlink{0000-0003-2552-0427}}
\affiliation{Department of Matter Science, University of Tebessa, 12000 Tebessa, Algeria}

\author{Nosratollah Jafari}
\email{nosrat.jafari@fai.kz}
\thanks{\orcidlink{0009-0008-3311-0480}}
\affiliation{Fesenkov Astrophysical Institute, 050020 Almaty, Kazakhstan}
\affiliation{Al-Farabi Kazakh National University, Al-Farabi av. 71, 050040 Almaty, Kazakhstan}
\affiliation{Center for Theoretical Physics, Khazar University, 41 Mehseti Street, AZ1096 Baku, Azerbaijan}

\date{\today}

\begin{abstract}
\secondrev{We study the Klein--Gordon oscillator in a doubly special relativity framework defined by a linear--fractional deformation of the mass-shell relation. In $(1+1)$ dimensions, the deformation is classified by the timelike, spacelike, or lightlike character of a fixed covector $a_\mu$. Using a reverted-product nonminimal coupling, we derive exact energy spectra and stationary states for all three sectors. The timelike and lightlike realizations are isospectral to each other and produce a common Planck-suppressed displacement of the particle and antiparticle branches, thereby breaking the exact $E\leftrightarrow -E$ symmetry. The spacelike realization leaves the spectrum unchanged but yields a non-Hermitian, $\mathcal{PT}$-symmetric spatial operator; an explicit similarity transformation maps it to the Hermitian Klein--Gordon oscillator. We also compare the first-power linear--fractional deformation with the Magueijo--Smolin model and show that the denominator power controls the magnitude of the spectral displacement. These results isolate how the causal class and functional form of the deformation affect relativistic oscillator spectra and states.}
\end{abstract}


\maketitle

\section{Introduction}

The possibility that special-relativistic kinematics may be modified near an observer-independent high-energy scale
$\Ep$ (equivalently, a short length scale $\lp=1/\Ep$ in units $\hbar=c=1$) is a recurring theme in quantum-gravity
phenomenology. Doubly special relativity (DSR) implements this idea while preserving the relativity principle by realizing
Lorentz transformations \emph{nonlinearly} on momentum space, which induces modified dispersion relations (MDRs)
\cite{AmelinoCamelia2001PLB,AmelinoCamelia2002IJMPD,AmelinoCamelia2002Open,MagueijoSmolin2002PRL,JudesVisser2003PRD,AmelinoCamelia2010Symmetry,Hossenfelder2010PRL}.
A complementary line of development connects DSR with $\kappa$-deformations of the Poincar\'e symmetry and with noncommutative
spacetime structures \cite{LukierskiRueggNowickiTolstoy1991PLB,MajidRuegg1994PLB,KowalskiGlikmanNowak2003IJMPD}.
Constraints on deviations from Lorentz-invariant propagation are typically very tight, yet the MDR framework remains a
useful theoretical laboratory and continues to motivate precision searches \cite{Mattingly2005LRR,Liberati2013CQG,AmelinoCamelia2013LRR,Hossenfelder2013LRR}.

\medskip
Exactly solvable relativistic bound-state systems provide a controlled setting in which to track how a deformed mass-shell invariant affects
(i) the spectrum, (ii) degeneracy patterns, and (iii) the analytic structure of the eigenfunctions. This is particularly useful
in DSR, because distinct nonlinear realizations may agree to leading order in $1/\Ep$ at the level of free-particle kinematics
while still producing qualitatively different operator representations once nonminimal couplings are introduced.
The KG oscillator is an especially convenient benchmark: it retains exact solvability, admits explicit closed-form
eigenfunctions, and keeps the two-branch (particle/antiparticle) structure of relativistic spectra transparent
\cite{BruceMinning1993NCimA,RaoKagali2008PhysScr,BakkeFurtado2015AOP}.
It is also closely related to the Dirac oscillator, which is introduced through a nonminimal substitution linear in both
position and momentum \cite{MoshinskySzczepaniak1989JPA,Quesne2017JPA}. In many settings, squaring the Dirac oscillator yields
KG-type oscillator sectors (plus spin-dependent couplings), which further motivates the KG oscillator as a clean starting
point for studies of deformed kinematics, including minimal-length and Planck-scale-deformed realizations \cite{KempfManganoMann1995PRD,Hossenfelder2013LRR}.

\medskip
In this work we solve the KG oscillator subject to a family of \emph{linear--fractional} deformed mass-shell-invariant constraints.
The deformation is generated by a single constant covector $a_\mu$, whose causal character leads to three inequivalent
DSR geometries (timelike, spacelike, lightlike). Importantly, although the free-particle dispersion relations differ only by
which momentum component appears in the deformation denominator, the oscillator implementation makes the differences
manifest: timelike/lightlike deformations displace the spectrum through a linear-in-$E$ term, whereas the spacelike case
produces a non-Hermitian (but $\mathcal{PT}$-symmetric) operator with complex-shifted eigenfunctions.

\medskip
\paragraph{Physical relevance of particle--antiparticle asymmetry.}
The breaking of the exact $E\leftrightarrow -E$ symmetry obtained in the timelike and lightlike geometries is one of the key
results of this paper, and it carries a physical content that goes beyond the DSR setting itself. In ordinary relativistic
quantum mechanics, the antisymmetry $E_{n,-}=-E_{n,+}$ is a direct consequence of the quadratic Casimir $p^\mu p_\mu+m^2=0$;
any departure from it must therefore be traced back either to non-quadratic dispersion or to non-Hermitian/non-self-conjugate
extensions of the dynamics. The linear-in-$E$ contribution induced by the linear--fractional deformation provides a clean,
exactly solvable example of the former mechanism: both branches are shifted by the same Planck-suppressed amount, so that
$E_{n,+}+E_{n,-}=-m^2/\Ep\neq 0$. This makes the system a useful theoretical analogue for situations in which particle and antiparticle branches are no longer
related by a simple sign reversal, including effective Lorentz-violation parametrizations and non-Hermitian deformations of
relativistic quantum mechanics with $\mathcal{PT}$ symmetry. The statement should not be read as a direct claim of observable
CPT violation: the leading common displacement cancels from transition energies inside a single branch, while its clean
signature is the nonzero branch sum and the comparison between positive- and negative-energy sectors. Even when the absolute
size of the shift is far below current experimental precision, the structure of the deformation discriminates between distinct
DSR realizations and is therefore a meaningful diagnostic for candidate quantum-gravity-inspired kinematics.

\paragraph{Context: pseudo-Hermiticity and $\mathcal{PT}$ symmetry.}
The non-Hermitian operator that emerges in the spacelike sector belongs to a class of imaginary-shifted oscillators that has been
studied extensively in non-Hermitian quantum mechanics, beginning with the seminal works of Bender and Boettcher on
$\mathcal{PT}$-symmetric Hamiltonians \cite{BenderBoettcher1998PRL,BenderBrodyJones2002PRL,Bender2007RPP} and the systematic pseudo-Hermitian formulation of
Mostafazadeh, together with the earlier quasi-Hermitian framework \cite{ScholtzGeyerHahne1992AP,Mostafazadeh2002JMP,Mostafazadeh2002JMP2,Mostafazadeh2002JMP3,Mostafazadeh2003JMP}. Closely related similarity transformations have been used to
diagonalize the Swanson model and its generalizations \cite{Swanson2004JMP,JonesMateo2005PRD}, and they appear in more recent
constructions such as $\mathcal{PT}$-regularization of conformal mechanics and the conformal bridge transformation, where a
similar non-unitary intertwiner maps a non-Hermitian operator to a Hermitian counterpart \cite{InzunzaPlyushchay2021JHEP,InzunzaPlyushchayWipf2020PRD}.
The complex displacement and momentum shift used in Sec.~4 are therefore standard tools in this framework, and related $\mathcal{PT}$-symmetric KG-oscillator constructions have also been discussed in the literature \cite{Cheng2011IJTP}. The novelty here is
not the technique itself but its application: it allows us to show that the spacelike linear--fractional DSR deformation
produces a fully consistent quantum theory with a real spectrum and a well-defined $\eta$-inner product, despite the explicitly
non-Hermitian appearance of the spatial operator.

\medskip
\noindent\textbf{Objectives and outline.}
The objectives of this paper are threefold.
First, we formulate and solve the Klein--Gordon oscillator under a linear--fractional (first-power) DSR deformation of the
mass-shell invariant generated by the nonlinear map introduced in Sec.~2, and we provide exact closed-form expressions for both
energy branches in the three inequivalent geometric realizations (timelike, spacelike, and lightlike).
Second, we clarify the qualitative differences between these geometries at the operator level: we show that the timelike and
lightlike choices induce an additive Planck-suppressed displacement of the spectrum and break the exact particle--antiparticle
symmetry through a term linear in $E$, whereas the spacelike choice is exactly isospectral but leads to complex-shifted
eigenfunctions and a non-Hermitian representation of the spatial operator.
Third, we provide a constructive $\mathcal{PT}$-symmetric and pseudo-Hermitian treatment of the spacelike sector (and of the
wavefunction sector of the lightlike geometry), exhibiting the explicit similarity transformation to a Hermitian oscillator,
the associated metric operator, and the resulting orthogonality structure.

In addition, we carry out a systematic comparison with the Magueijo--Smolin DSR model, emphasizing how the power of the
denominator in the deformed invariant controls the magnitude and the functional form of the Planck-suppressed spectral shifts at
fixed $m/\Ep$. To make the parameter dependence transparent we introduce a dimensionless formulation and present
representative plots illustrating the dependence on the deformation ratio $m/\Ep$, the oscillator strength $\omega/m$, and the
excitation level $n$.

The paper is organized as follows. Section~2 introduces the linear--fractional mass-shell-invariant deformation and its classification
into timelike, spacelike, and lightlike geometries. Section~3 describes the reverted-product realization of the KG oscillator
and the operator implementation of the deformed dispersion relations. Section~4 presents the exact spectra and eigenfunctions
in the three geometries, including the $\mathcal{PT}$-symmetric construction. Section~5 compares the results with
Magueijo--Smolin DSR. Section~6 develops the dimensionless analysis and the numerical illustrations. Section~7 presents the
outlook, including a brief discussion of possible extensions toward spinorial oscillator models, and Sec.~8 concludes.
An appendix collects the metric-operator and $\eta$-orthogonality details for the spacelike and lightlike sectors.

\medskip
We work in $(1+1)$ dimensions with Minkowski signature $\eta_{\mu\nu}=\mathrm{diag}(-1,+1)$, so that
$\eta_{\mu\nu}p^\mu p^\nu=-E^2+p^2$ and the special-relativistic dispersion relation reads $E^2-p^2=m^2$.

\section{Linear--Fractional Deformed Mass-Shell Invariant and Geometric Classification}

\subsection{Nonlinear map and the first-power mass-shell invariant}
Let $p^\mu$ denote physical momenta and let $\pi^\mu$ be auxiliary variables transforming linearly under Lorentz
transformations, $\pi'^{\mu}=\Lambda^\mu{}_\nu\pi^\nu$. Consider the nonlinear map
\begin{equation}
\pi^\mu=\frac{p^\mu}{\sqrt{\,1+\lp\, a_\alpha p^\alpha\,}},
\label{eq:map}
\end{equation}
where $a_\mu$ is a constant covector and $\lp=1/\Ep$.
Imposing the standard auxiliary mass-shell condition
\begin{equation}
\eta_{\mu\nu}\pi^\mu\pi^\nu=-m^2
\end{equation}
and using \eqref{eq:map} yields the \emph{linear--fractional} (first-power) deformed mass-shell invariant:
\begin{equation}
\frac{\eta_{\mu\nu}p^\mu p^\nu}{1+\lp\,a_\alpha p^\alpha}=-m^2.
\label{eq:CasimirGeneral}
\end{equation}

\paragraph{Induced nonlinear Lorentz action and invariance.}
Because $\pi^\mu$ transforms linearly, the auxiliary quadratic form $\eta_{\mu\nu}\pi^\mu\pi^\nu$ is manifestly invariant under
ordinary Lorentz transformations. The physical momentum $p^\mu$ inherits a nonlinear realization of the Lorentz group through
the map $F:p^\mu\mapsto\pi^\mu$ defined by \eqref{eq:map}: explicitly,
\begin{equation}
p'^\mu = F^{-1}\!\left(\Lambda^\mu{}_\nu\,F(p)^\nu\right).
\label{eq:nonlinearLorentz}
\end{equation}
{\color{blue}The inverse map is obtained by setting $p^\mu=s\pi^\mu$ and defining $A\equiv a\cdot\pi$. Substitution into
\eqref{eq:map} gives, for $\pi^\mu\neq0$,
\[
\pi^\mu=\frac{s\pi^\mu}{\sqrt{1+\lp sA}}
\quad\Longrightarrow\quad
s^2=1+\lp sA,
\]
and therefore
\[
s^2-\lp A s-1=0.
\]
For $\pi^\mu=0$, the same result is obtained by continuity. Solving this quadratic equation and selecting the branch
that is continuously connected to the undeformed limit gives
\begin{equation}
s=\frac{\lp\,(a\cdot\pi)+\sqrt{4+\lp^2(a\cdot\pi)^2}}{2},
\qquad
p^\mu=F^{-1}(\pi)^\mu
=\frac{\lp\,(a\cdot\pi)+\sqrt{4+\lp^2(a\cdot\pi)^2}}{2}\,\pi^\mu.
\label{eq:inversemap}
\end{equation}
Indeed, the other root tends to $-1$ as $\lp\to0$, whereas the branch in \eqref{eq:inversemap} satisfies $s\to1$ and
hence $F^{-1}(\pi)^\mu\to\pi^\mu$.

The scalar inverse follows directly by contracting $p^\mu=s\pi^\mu$ with $a_\mu$:
\[
a\cdot p=sA.
\]
Using the quadratic relation $s^2=1+\lp sA$ then gives
\[
\lp(a\cdot p)=\lp sA=s^2-1.
\]
Substitution of the positive solution for $s$ yields
\begin{equation}
a\cdot p=\frac{1}{\lp}\left[\left(\frac{\lp(a\cdot\pi)+\sqrt{4+\lp^2(a\cdot\pi)^2}}{2}\right)^2-1\right].
\label{eq:inversemapScalar}
\end{equation}
The equivalence of this expression with $a\cdot p=s(a\cdot\pi)$ can also be checked directly. With
$A=a\cdot\pi$,
\[
\begin{aligned}
s^2-1
&=\frac{\lp^2A^2+\lp A\sqrt{4+\lp^2A^2}}{2}\\
&=\lp A\,\frac{\lp A+\sqrt{4+\lp^2A^2}}{2}
=\lp sA.
\end{aligned}
\]
Thus the right-hand side of \eqref{eq:inversemapScalar} is exactly $sA=a\cdot p$. Finally, the small-$\lp$
expansion
\[
s=1+\frac{\lp A}{2}+\frac{\lp^2A^2}{8}+O\!\left((\lp A)^3\right),
\qquad
a\cdot p=A+\frac{\lp A^2}{2}+\frac{\lp^2A^3}{8}+O(\lp^3A^4),
\]
shows explicitly that the apparent $1/\lp$ singularity in \eqref{eq:inversemapScalar} is removable and that the inverse
map reduces smoothly to the identity in the undeformed limit.}
Under this induced nonlinear transformation, the combination $\eta_{\mu\nu}\pi^\mu\pi^\nu$, and therefore the linear--fractional
quantity $\eta_{\mu\nu}p^\mu p^\nu/(1+\lp\,a_\alpha p^\alpha)$, is preserved. This is the precise sense in which
\eqref{eq:CasimirGeneral} is an invariant of the nonlinearly realized Lorentz symmetry of the present DSR scheme, in agreement
with the general construction of DSR models based on nonlinear maps in momentum space \cite{JudesVisser2003PRD,MagueijoSmolin2002PRL}.
We do not postulate here a separate quantum-group Hopf algebra or a unique set of deformed Poincar\'e commutators. Instead,
the calculation uses the standard DSR construction in which the ordinary Lorentz action on the auxiliary momentum space is
pulled back to a nonlinear action on the physical momenta. To avoid possible ambiguity with the strict notion of a Casimir
element of a Lie algebra, we refer to \eqref{eq:CasimirGeneral} as a \emph{linear--fractional deformed mass-shell invariant}
(or equivalently as a deformed dispersion relation), and reserve the word ``Casimir'' only for the qualified sense associated
with this nonlinear realization.

\paragraph{Equivalent MDR form and small-$\lp$ expansion.}
Multiplying \eqref{eq:CasimirGeneral} by the denominator gives
\begin{equation}
\eta_{\mu\nu}p^\mu p^\nu + m^2
=
-\,m^2\lp\,(a\cdot p),
\label{eq:MDRlinear}
\end{equation}
which makes explicit that the first-power deformation introduces a \emph{linear} correction in the momentum component
selected by $a_\mu$. In the perturbative regime $\lp\,|a\cdot p|\ll 1$ the deformation is Planck-suppressed and the map reduces
to the identity at leading order,
\begin{equation}
\pi^\mu = p^\mu \left[1-\frac{\lp}{2}(a\cdot p)+O(\lp^2)\right].
\end{equation}
Although \eqref{eq:MDRlinear} is algebraically simple, its geometric content is richer than this expansion suggests, because the
different causal classes of $a_\mu$ cannot be mapped into each other by Lorentz transformations.

\paragraph{Physical domain.}
The map \eqref{eq:map} is well defined in the region where $1+\lp\,a\cdot p>0$. In timelike realizations this is commonly
interpreted as restricting the energy away from a pole at $E\simeq \Ep$, whereas in spacelike or lightlike realizations the
restriction involves the spatial momentum or light-cone combinations. In the bound-state setting we choose parameters such that
all plotted levels remain in this regular domain, so that the deformation can be treated consistently within the stationary
ansatz used below.

\subsection{Timelike, spacelike, and lightlike choices of $a_\mu$}
In $(1+1)$ dimensions one may choose representatives so that the denominator depends on $E$, $p$, or $E+p$:
\begin{align}
\text{timelike:}\quad & a_\mu=(-1,0), \qquad a_\alpha p^\alpha=-E, \\
\text{spacelike:}\quad & a_\mu=(0,-1), \qquad a_\alpha p^\alpha=-p, \\
\text{lightlike:}\quad & a_\mu=(-1,-1), \qquad a_\alpha p^\alpha=-(E+p).
\end{align}
Then \eqref{eq:CasimirGeneral} yields the three inequivalent first-power MDRs
\begin{align}
\frac{E^2-p^2}{1-E/\Ep}&=m^2, \label{eq:CasimirTL}\\
\frac{E^2-p^2}{1-p/\Ep}&=m^2, \label{eq:CasimirSL}\\
\frac{E^2-p^2}{1-(E+p)/\Ep}&=m^2. \label{eq:CasimirLL}
\end{align}
These reduce to $E^2-p^2=m^2$ as $\Ep\to\infty$. The causal classification is closely aligned with how time-, space-, and
lightlike $\kappa$-deformations arise in quantum-group approaches to deformed kinematics
\cite{LukierskiRueggNowickiTolstoy1991PLB,MajidRuegg1994PLB,KowalskiGlikmanNowak2002PLB,KowalskiGlikmanNowak2003CQG}.

\paragraph{Geometric meaning.}
Because $a_\mu$ is a fixed covector, it selects a preferred direction in momentum space \emph{only} in the sense that the
nonlinear realization is written with respect to that direction. The representative choices of $a_\mu$ label inequivalent causal
classes of the deformation. Within each class, the nonlinear Lorentz action is defined by the pullback through $F$; no independent
Hopf-algebra structure is assumed here. The relativity principle is maintained, because the auxiliary variables $\pi^\mu$ transform
linearly and the entire deformation is encoded in the nonlinear map between $p^\mu$ and $\pi^\mu$.
In $(1+1)$ dimensions the three causal classes (timelike, spacelike, lightlike) exhaust the inequivalent cases, which is why the
oscillator problem naturally splits into three sectors with distinct spectral and operator-theoretic behavior.

\subsection{Relation to the Magueijo--Smolin (MS) DSR invariant}
\label{sec:compareMS}
In the Magueijo--Smolin realization (often called DSR2), the invariant can be written as
\begin{equation}
\frac{\eta_{\mu\nu}p^\mu p^\nu}{(1-\lp\,p_0)^2}=-m^2,
\end{equation}
which in $(1+1)$ dimensions becomes
\begin{equation}
\frac{E^2-p^2}{(1-E/\Ep)^2}=m^2.
\label{eq:MSdispersion}
\end{equation}
The key structural distinction with \eqref{eq:CasimirTL} is the \emph{power} of the denominator. This changes the way the pole
at $E=\Ep$ enters the mass shell, and consequently changes the scale of the Planck-suppressed spectral displacements found in
bound-state problems. Expanding at large $\Ep$, one finds
\begin{align}
\text{first-power:}\quad & E^2-p^2 = m^2\left(1-\frac{E}{\Ep}\right)+O(\Ep^{-2}),\\
\text{MS (squared):}\quad & E^2-p^2 = m^2\left(1-\frac{2E}{\Ep}\right)+O(\Ep^{-2}),
\end{align}
which anticipates the fact that, at fixed $m/\Ep$, the MS model produces a leading displacement approximately twice as large as
the first-power timelike or lightlike model.

\section{Klein--Gordon Oscillator and Operator Realization}

\subsection{Reverted-product nonminimal coupling}
The KG oscillator is introduced by a nonminimal substitution in the spatial momentum sector. A particularly transparent
ordering is the reverted-product operator
\begin{equation}
\hat{P}^2_{\mathrm{rev}}
=
(\hat{p}+\ii m\omega x)(\hat{p}-\ii m\omega x)
=
\hat{p}^2+m^2\omega^2x^2-m\omega,
\label{eq:Prev}
\end{equation}
where $\hat{p}=-\ii\,\dd/\dd x$ and $[x,\hat{p}]=\ii$.

The eigenstructure of \eqref{eq:Prev} is most economically obtained from the standard ladder operators. Defining
\begin{equation}
a=\frac{1}{\sqrt{2m\omega}}\left(m\omega x+\ii \hat{p}\right),
\qquad
a^\dagger=\frac{1}{\sqrt{2m\omega}}\left(m\omega x-\ii \hat{p}\right),
\qquad [a,a^\dagger]=1,
\end{equation}
a short computation gives
\begin{equation}
\hat{P}^{\,2}_{\mathrm{rev}}=2m\omega\,a^\dagger a,
\end{equation}
so $\hat{P}^{\,2}_{\mathrm{rev}}$ is, up to the multiplicative factor $2m\omega$, the harmonic-oscillator number operator. Its
spectrum is therefore 
\begin{equation}
\hat{P}^2_{\mathrm{rev}}\psi_n = \lambda_n\psi_n,\qquad
\lambda_n=2n\,m\omega,\qquad n=0,1,2,\ldots
\label{eq:lambda}
\end{equation}
with $\lambda_0=0$, and the eigenfunctions coincide with those of the ordinary one-dimensional harmonic oscillator.
We use the stationary ansatz $\Psi(t,x)=e^{-\ii Et}\psi(x)$, so that $\hat{E}\to E$ throughout the spectral analysis below.

\paragraph{Remark on ordering.}
The reverted-product choice \eqref{eq:Prev} is not merely cosmetic: it preserves exact solvability and produces the clean
eigenvalue reduction \eqref{eq:lambda} with vanishing ground-state energy $\lambda_0=0$. Other orderings differ by constant shifts
(effectively $n\to n+1$) and may obscure direct quantitative comparisons between distinct MDR models, especially when
Planck-suppressed additive shifts are also present.

\subsection{Promoting the deformed mass-shell relation to an operator constraint}
In each geometry we promote the dispersion relation to an operator equation by substituting $p^2\to\hat{P}^2_{\mathrm{rev}}$
and treating $E$ as a spectral parameter in the stationary sector. Where a linear $p$ appears (spacelike/lightlike), we
implement the oscillator consistently by using the same nonminimal structure in the linear term:
\begin{equation}
p \ \longrightarrow\ (\hat{p}+\ii m\omega x),
\label{eq:linpRule}
\end{equation}
which is the natural companion to \eqref{eq:Prev}. This choice keeps the deformation aligned with the oscillator's
creation/annihilation structure and is crucial for enabling the explicit similarity transformation used below.

\paragraph{On operator ordering and physical observables.}
In deformed kinematics, different operator realizations may be related by similarity transformations or may correspond to
different prescriptions for coupling the MDR to interactions. Our aim here is to adopt a minimal and analytically tractable
realization that (i) reduces to the usual KG oscillator as $\Ep\to\infty$, (ii) preserves exact solvability, and (iii) makes
the causal distinction of $a_\mu$ operational at the spectral and wavefunction level. The pseudo-Hermitian construction in
Sec.~4 further guarantees a consistent inner-product structure whenever non-Hermitian operators occur.

\section{Exact Eigensolutions in the Three Geometries}

\subsection{Timelike geometry}
From \eqref{eq:CasimirTL} with $p^2\to\hat{P}_{\mathrm{rev}}^2$:
\begin{equation}
\frac{E^2-\hat{P}_{\mathrm{rev}}^2}{1-E/\Ep}=m^2
\quad\Longrightarrow\quad
E^2+\frac{m^2}{\Ep}E-\left(\hat{P}_{\mathrm{rev}}^2+m^2\right)=0.
\label{eq:TLop}
\end{equation}
Acting on $\psi_n$ and using \eqref{eq:lambda} yields
\begin{equation}
E^2+\frac{m^2}{\Ep}E-\left(m^2+2n\,m\omega\right)=0,
\label{eq:TLquad}
\end{equation}
so that the two branches are
\begin{equation}
E_{n,\pm}^{\mathrm{TL}}
=
-\frac{m^2}{2\Ep}\pm\sqrt{m^2+2n\,m\omega+\frac{m^4}{4\Ep^2}}.
\label{eq:ETLbranches}
\end{equation}

\paragraph{Branch structure and displacement.}
Equation \eqref{eq:ETLbranches} makes two basic effects explicit:
(i) the spectrum remains real and discrete, and (ii) both branches are shifted by the same additive amount
$-m^2/(2\Ep)$, breaking the exact particle--antiparticle symmetry $E_{n,-}=-E_{n,+}$.
A simple diagnostic is the branch sum,
\begin{equation}
E_{n,+}^{\mathrm{TL}}+E_{n,-}^{\mathrm{TL}}=-\frac{m^2}{\Ep},
\end{equation}
which vanishes in the undeformed oscillator but is fixed here by the deformation scale.

\paragraph{Physical interpretation.}
The nonzero branch sum is the operator-level manifestation of the linear-in-$E$ term in the deformed dispersion
relation~\eqref{eq:MDRlinear}. In the special-relativistic limit $\Ep\to\infty$ the dispersion relation is even in $E$, so the two
roots of the mass shell are antipodal and the bound-state spectrum is automatically symmetric. Once a linear-in-$E$ contribution is
present, the two roots become asymmetric by a common Planck-suppressed amount, and the bound-state spectrum inherits this rigid
displacement at every excitation level. Equivalently, particles and antiparticles ``see'' slightly different effective rest masses,
even though no source of CPT violation has been added by hand. This makes the timelike (and lightlike) sector of the
linear--fractional model a controlled analogue, within solvable bound-state quantum mechanics, of mass-asymmetry effects discussed
more generally in Lorentz-invariance-violation phenomenology and in non-Hermitian extensions of relativistic field theory. The
quantitative size of the effect is, of course, far below current experimental sensitivity for realistic Planck scales; moreover,
because the leading displacement is common to all levels in a given branch, ordinary intrabranch transition energies are not
shifted at first order. Nevertheless, the nonzero branch sum and the comparison of the two branches unambiguously distinguish
the present DSR realization from the spacelike one (which is isospectral) and from the Magueijo--Smolin model (where the shift
is twice as large at fixed $m/\Ep$, see Sec.~\ref{sec:MScomparison}).

\paragraph{Perturbative regime and admissible levels.}
In the regime $\Ep\gg m$ one finds
\begin{equation}
E_{n,\pm}^{\mathrm{TL}}
=
\pm\sqrt{m^2+2n\,m\omega}\;-\;\frac{m^2}{2\Ep}+O(\Ep^{-2}).
\label{eq:ETLasympt}
\end{equation}
The timelike model contains a pole at $E=\Ep$. A conservative condition for remaining in the regular domain is
$E_{n,+}/\Ep<1$, which (at leading order) suggests
\begin{equation}
n \lesssim \frac{\Ep^2-m^2}{2m\omega}.
\label{eq:nmaxTimelike}
\end{equation}
In practical illustrative plots (Sec.~6) we choose parameters so that all plotted positive-branch levels satisfy this
constraint comfortably.

\subsubsection{Energy reparametrization viewpoint for timelike/lightlike sectors}
\label{sec:energyreparam}
A useful way to interpret the linear term in $E$ is to complete the square in \eqref{eq:TLquad}. Writing
\begin{equation}
\left(E+\frac{m^2}{2\Ep}\right)^2
=
m^2+2n\,m\omega+\frac{m^4}{4\Ep^2},
\end{equation}
we introduce the shifted variable
\begin{equation}
\widetilde{E}:=E+\frac{m^2}{2\Ep}.
\end{equation}
In terms of $\widetilde{E}$ the spectrum becomes symmetric:
\begin{equation}
\widetilde{E}_{n,\pm}=\pm\sqrt{m^2+2n\,m\omega+\frac{m^4}{4\Ep^2}},
\end{equation}
while the \emph{physical} energies $E_{n,\pm}=\widetilde{E}_{n,\pm}-m^2/(2\Ep)$ inherit an overall displacement.
Thus, one may regard the timelike/lightlike deformation as implementing a universal Planck-suppressed reparametrization
of the energy origin: the two branches remain opposite in $\widetilde{E}$, but are both shifted in $E$.
This viewpoint is convenient when comparing deformations: at leading order, the first-power deformation yields
$\Delta E\simeq -m^2/(2\Ep)$, whereas MS yields $\Delta E\simeq -m^2/\Ep$ (cf.\ Sec.~\ref{sec:MScomparison}).

\paragraph{Wavefunctions.}
The spatial eigenproblem remains the standard reverted-product oscillator; thus one may take
$\psi_n^{\mathrm{TL}}(x)=\phi_n(x)$, where $\phi_n(x)$ is the standard one-dimensional harmonic-oscillator
eigenfunction (Hermite--Gaussian). The deformation therefore
affects the timelike sector \emph{only} through the energy eigenvalues, not through spatial localization properties at the
level of $\psi(x)$.

\subsection{Spacelike geometry}
Starting from \eqref{eq:CasimirSL} and implementing the oscillator in the linear term according to \eqref{eq:linpRule} gives
\begin{equation}
E^2\psi
=
\left[
\hat{P}_{\mathrm{rev}}^2+m^2-\frac{m^2}{\Ep}\,(\hat{p}+\ii m\omega x)
\right]\psi.
\label{eq:SLop}
\end{equation}
Using \eqref{eq:Prev}, one finds
\begin{equation}
\begin{aligned}
E^2\psi
&=
\biggl[
\left(\hat{p}^2-\frac{m^2}{\Ep}\hat{p}\right)
+
\left(m^2\omega^2x^2-\ii\frac{m^3\omega}{\Ep}x\right)
+m^2-m\omega
\biggr]\psi.
\end{aligned}
\label{eq:SLexpanded}
\end{equation}
While \eqref{eq:SLexpanded} is non-Hermitian in the standard $L^2(\mathbb{R})$ inner product, it remains exactly solvable
and admits a consistent $\eta$-inner product (pseudo-Hermiticity).

\paragraph{Completing squares and isospectrality.}
Completing the square in momentum and position gives
\begin{align}
\hat{p}^2-\frac{m^2}{\Ep}\hat{p}
&=
\left(\hat{p}-\frac{m^2}{2\Ep}\right)^2-\frac{m^4}{4\Ep^2},\\
m^2\omega^2x^2-\ii\frac{m^3\omega}{\Ep}x
&=
m^2\omega^2\left(x-\ii\frac{m}{2\omega\Ep}\right)^2+\frac{m^4}{4\Ep^2},
\end{align}
so the constants cancel and one obtains
\begin{equation}
E^2\psi
=
\left[
(\hat{p}')^2+m^2\omega^2(x')^2+m^2-m\omega
\right]\psi,
\qquad
\hat{p}'=\hat{p}-\frac{m^2}{2\Ep},\quad
x'=x-\ii\frac{m}{2\omega\Ep}.
\label{eq:SLshifted}
\end{equation}
Hence
\begin{equation}
E^2 = m^2+2n\,m\omega,
\qquad\Rightarrow\qquad
E_{n,\pm}^{\mathrm{SL}}=\pm\sqrt{m^2+2n\,m\omega}.
\label{eq:ESLbranches}
\end{equation}
The spacelike deformation is therefore \emph{isospectral}: it changes eigenfunctions but not eigenvalues. In particular,
the antisymmetry $E_{n,-}^{\mathrm{SL}}=-E_{n,+}^{\mathrm{SL}}$ is preserved, so that the spacelike sector does \emph{not} break
the exact particle--antiparticle symmetry at the spectral level. Its physical content is shifted from the eigenvalues to the
eigenfunctions and to the non-Hermitian, $\mathcal{PT}$-symmetric structure of the spatial operator, in sharp contrast with the
timelike and lightlike geometries discussed above.

\subsubsection{Explicit eigenfunctions, $\mathcal{PT}$ symmetry, and pseudo-Hermiticity}
\label{sec:PTconstruction}

\paragraph{Non-Hermitian operator and similarity map.}
Denote the right-hand side of \eqref{eq:SLexpanded} by $H$, so that the spacelike eigenproblem reads $H\psi=E^2\psi$ with
\begin{equation}
H=\hat{p}^2-\frac{m^2}{\Ep}\,\hat{p}+m^2\omega^2 x^2-\ii\,\frac{m^3\omega}{\Ep}\,x+m^2-m\omega.
\label{eq:HSL}
\end{equation}
The operator $H$ is non-Hermitian on $L^2(\mathbb{R})$ because of the imaginary linear term in $x$, but it can be mapped onto a
manifestly Hermitian oscillator operator by a single similarity transformation.

Define the deformation-induced shifts
\begin{equation}
\delta=\frac{m}{2\omega\Ep},\qquad \kappa=\frac{m^2}{2\Ep},
\label{eq:kappadelta}
\end{equation}
and introduce the (generally non-unitary) similarity operator
\begin{equation}
\mathcal{S}=e^{-\delta \hat{p}}\,e^{-\ii\kappa x},
\label{eq:similarityS}
\end{equation}
whose action on the canonical variables is
\begin{equation}
\mathcal{S}^{-1}x\,\mathcal{S}=x-\ii\delta,
\qquad
\mathcal{S}^{-1}\hat{p}\,\mathcal{S}=\hat{p}-\kappa.
\label{eq:Saction}
\end{equation}
Inserting \eqref{eq:Saction} in \eqref{eq:HSL} and collecting the resulting cross-terms, the explicit Hermitian
operator $h$ similar to $H$ is
\begin{equation}
\boxed{\;h=\hat{p}^2+m^2\omega^2 x^2+m^2-m\omega\;,
\qquad H=\mathcal{S}^{-1}h\,\mathcal{S},\qquad h^\dagger=h.\;}
\label{eq:hExplicit}
\end{equation}
This is exactly the undeformed Klein--Gordon oscillator squared-energy operator: the spacelike linear--fractional
deformation acts at the level of the wavefunctions through the non-unitary intertwiner $\mathcal{S}$, but leaves the spectrum
of the underlying Hermitian operator unchanged. Equation~\eqref{eq:hExplicit} therefore makes the isospectrality
$E_{n,\pm}^{\mathrm{SL}}=\pm\sqrt{m^2+2nm\omega}$ manifest at the operator level and provides the explicit Hermitian
counterpart whose eigenfunctions are the standard Hermite--Gaussian states.

\paragraph{Eigenfunctions.}
Let $\phi_n$ denote the standard one-dimensional harmonic-oscillator eigenfunction,
\begin{equation}
\phi_n(x)=\mathcal{N}_n\exp\!\left(-\frac{m\omega}{2}x^2\right)H_n\!\left(\sqrt{m\omega}\,x\right),
\qquad
\mathcal{N}_n=\left(\frac{m\omega}{\pi}\right)^{1/4}\frac{1}{\sqrt{2^n\,n!}}\,,
\end{equation}
with $\mathcal{N}_n$ the standard $L^2(\mathbb{R})$ normalization constant. A corresponding eigenfunction of \eqref{eq:SLexpanded} is then
\begin{equation}
\psi_n^{\mathrm{SL}}(x)=e^{\ii\kappa x}\,\phi_n(x-\ii\delta).
\label{eq:psiform}
\end{equation}
The spatial profile is therefore a complex translation of the Hermite--Gaussian, accompanied by a plane-wave phase.

\paragraph{Why the spectrum remains real.}
With parity $\mathcal{P}: x\to -x,\ \hat{p}\to -\hat{p}$ and time reversal
$\mathcal{T}: \ii\to -\ii,\ \hat{p}\to -\hat{p}$, the combination $-\ii x$ is $\mathcal{PT}$-even. Consequently the
imaginary linear term in $x$ in \eqref{eq:SLexpanded} is $\mathcal{PT}$ symmetric. More strongly, the similarity map
\eqref{eq:similarityS} yields pseudo-Hermiticity and the associated metric operator and orthogonality relations,
summarized in Appendix~\ref{app:metric} \cite{BenderBoettcher1998PRL,Mostafazadeh2002JMP,Mostafazadeh2003JMP}. Similarity
transformations of this type are standard in the analysis of imaginary-shifted oscillators and underlie, in particular, the
diagonalization of the Swanson model and related constructions \cite{Swanson2004JMP,JonesMateo2005PRD,InzunzaPlyushchay2021JHEP}.

\paragraph{Physical inner product and observables (interpretive note).}
Because $H$ is non-Hermitian in the standard inner product, expectation values should be computed with the $\eta$-inner
product $(\cdot,\cdot)_\eta$ induced by the metric $\eta=\mathcal{S}^\dagger\mathcal{S}$. In this framework, the spectrum
is real and the eigenfunctions form an $\eta$-orthonormal basis, so the spacelike deformation defines a consistent quantum
theory despite the non-Hermitian appearance of \eqref{eq:SLexpanded}.

\subsection{Lightlike geometry}
From \eqref{eq:CasimirLL} one obtains, after operator promotion,
\begin{equation}
\begin{aligned}
E^2+\frac{m^2}{\Ep}E
-\left(\hat{P}_{\mathrm{rev}}^2+m^2\right)
+\frac{m^2}{\Ep}(\hat{p}+\ii m\omega x)
=0.
\end{aligned}
\label{eq:LLop}
\end{equation}
The last term has the same structure as in the spacelike sector and is removed by the same similarity map; consequently,
one again obtains \eqref{eq:TLquad} on oscillator eigenstates:
\begin{equation}
E_{n,\pm}^{\mathrm{LL}}
=
-\frac{m^2}{2\Ep}\pm\sqrt{m^2+2n\,m\omega+\frac{m^4}{4\Ep^2}}.
\label{eq:ELLbranches}
\end{equation}
Thus, the lightlike deformation shares the \emph{timelike} spectrum but inherits spacelike-type complex-shifted eigenfunctions:
\begin{equation}
\psi_n^{\mathrm{LL}}(x)=e^{\ii\kappa x}\,\phi_n(x-\ii\delta),
\end{equation}
with $\kappa,\delta$ as in \eqref{eq:kappadelta}. The energy-reparametrization interpretation of
Sec.~\ref{sec:energyreparam} applies equally to this sector.

\paragraph{Interpretation.}
The lightlike case can be viewed as combining (i) the timelike-type linear term in $E$ that displaces both branches and
(ii) the spacelike-type linear momentum contribution that produces a non-Hermitian but $\mathcal{PT}$-symmetric spatial
operator. This makes the lightlike realization a useful ``hybrid'' laboratory for disentangling which aspects of the
deformation affect spectral positions versus wavefunction structure.

\subsection{Summary of the three geometries}
For convenience, we collect in Table~\ref{tab:summary} the essential outputs of the three inequivalent first-power DSR
realizations in $(1+1)$ dimensions: the deformation denominator that specifies the MDR geometry and the corresponding
closed-form energy branches. The representative stationary eigenfunctions are given explicitly in the preceding subsections.
This summary makes transparent that
timelike and lightlike choices produce the same Planck-suppressed \emph{spectral displacement}, whereas the spacelike choice
is strictly \emph{isospectral} but modifies the states through a complex translation (with a consistent pseudo-Hermitian
$\eta$-inner product).
\begin{table*}[t]
\caption{Summary of the three inequivalent first-power DSR geometries in $(1+1)$ dimensions, with the
explicit closed-form energy branches written out.}
\label{tab:summary}
\centering
\begin{ruledtabular}
\begin{tabular}{@{}l l l@{}}
Geometry & Deformation denominator & Energy branches $E_{n,\pm}$ \\
\hline
\\[-0.6em]
Timelike   & $1-E/\Ep$       & $\displaystyle -\frac{m^2}{2\Ep}\pm\sqrt{\,m^2+2nm\omega+\frac{m^4}{4\Ep^{2}}\,}$ \\[0.9em]
Spacelike  & $1-p/\Ep$       & $\displaystyle \pm\sqrt{\,m^2+2nm\omega\,}$ \\[0.9em]
Lightlike  & $1-(E+p)/\Ep$   & $\displaystyle -\frac{m^2}{2\Ep}\pm\sqrt{\,m^2+2nm\omega+\frac{m^4}{4\Ep^{2}}\,}$ \\[0.4em]
\end{tabular}
\end{ruledtabular}
\end{table*}

\section{Comparison with Magueijo--Smolin DSR}
\label{sec:MScomparison}
For MS DSR, the dispersion relation is \eqref{eq:MSdispersion}. Implementing the oscillator via
$p^2\to\hat{P}^2_{\mathrm{rev}}$ and acting on $\psi_n$ gives
\begin{equation}
\frac{E^2-\lambda_n}{(1-E/\Ep)^2}=m^2,
\qquad \lambda_n=2n\,m\omega,
\end{equation}
which leads to
\begin{equation}
\left(1-\frac{m^2}{\Ep^2}\right)E^2+\frac{2m^2}{\Ep}E-(m^2+2n\,m\omega)=0.
\label{eq:MSquad}
\end{equation}
Solving this quadratic equation in $E$ gives
\begin{equation}
\begin{aligned}
E_{n,\pm}^{(\mathrm{MS})}
&=
\frac{1}{1-\frac{m^2}{\Ep^2}}
\biggl[
-\frac{m^2}{\Ep}
\pm
\sqrt{\,m^2+2n\,m\omega\!\left(1-\frac{m^2}{\Ep^2}\right)}
\biggr].
\end{aligned}
\label{eq:MSbranches}
\end{equation}
For $\Ep\gg m$,
\begin{equation}
E_{n,\pm}^{(\mathrm{MS})}
=
\pm\sqrt{m^2+2n\,m\omega}\;-\;\frac{m^2}{\Ep}+O(\Ep^{-2}).
\label{eq:MSasympt}
\end{equation}

\paragraph{Quantitative comparison and role of the denominator power.}
Both models break the exact $E\leftrightarrow -E$ symmetry by shifting the two branches by the same additive amount, but the
magnitude of the shift differs: MS gives a leading displacement that is twice as large as in the first-power timelike or
lightlike model \eqref{eq:ETLasympt}. Beyond leading order, MS introduces in addition the prefactor $(1-m^2/\Ep^2)^{-1}$, which
produces $O(\Ep^{-2})$ distortions; the MS deformation is therefore not reducible to a pure energy-origin shift once higher-order
terms are kept. This shows that the power of the denominator is a genuinely model-defining input in DSR-inspired bound-state
spectra.

\paragraph{Pole structure and admissible parameters.}
Both the first-power timelike model and MS contain a pole at $E=\Ep$. For phenomenologically realistic masses the ratio
$\epsilon=m/\Ep$ is tiny, so this pole sits far above the oscillator energies unless $\omega$ or $n$ are taken extremely large.
In the illustrative plots of Sec.~6 we nevertheless enforce a conservative ``below-pole'' criterion in order to avoid spurious
near-pole artefacts.

\section{Parameter Dependence and Numerical Illustration}
\label{sec:plots}

\subsection{Dimensionless form}
Introduce the dimensionless ratios
\begin{equation}
\Omega=\frac{\omega}{m},\qquad \epsilon=\frac{m}{\Ep},\qquad e_{n,\pm}=\frac{E_{n,\pm}}{m}.
\label{eq:dimensionless}
\end{equation}
The special-relativistic spectrum reads
\begin{equation}
e_{n,\pm}^{\mathrm{SR}}=\pm\sqrt{1+2n\Omega}.
\end{equation}
For the first-power mass-shell-invariant model:
\begin{align}
\text{spacelike:}\quad & e_{n,\pm}^{\mathrm{SL}}=\pm\sqrt{1+2n\Omega}, \\
\text{timelike/lightlike:}\quad &
e_{n,\pm}^{\mathrm{TL}}=e_{n,\pm}^{\mathrm{LL}}
=-\frac{\epsilon}{2}\pm\sqrt{1+2n\Omega+\frac{\epsilon^2}{4}}.
\end{align}
For MS:
\begin{equation}
e_{n,\pm}^{(\mathrm{MS})}
=
\frac{-\epsilon \pm \sqrt{1+2n\Omega(1-\epsilon^2)}}{1-\epsilon^2}.
\end{equation}

\paragraph{Interpretation of the dimensionless parameters.}
$\Omega=\omega/m$ measures the oscillator strength relative to the rest mass, while $\epsilon=m/\Ep$ controls the magnitude of
the DSR corrections. In realistic particle-physics settings $\epsilon\ll 1$, so the DSR-induced spectral displacement is tiny and
is essentially $n$-independent at leading order. In quantum-mechanical analogue contexts or in purely theoretical explorations,
on the other hand, larger values of $\epsilon$ may be used in order to visualize the structure of the deformation, while keeping
the interpretation of the Planck scale as an effective cutoff.

\subsection{Leading-order interpretation}
For $\epsilon\ll 1$ one finds
\begin{align}
e_{n,+}^{\mathrm{TL}}-e_{n,+}^{\mathrm{SR}} &= -\frac{\epsilon}{2}+O(\epsilon^2),\\
e_{n,+}^{(\mathrm{MS})}-e_{n,+}^{\mathrm{SR}} &= -\epsilon+O(\epsilon^2),
\end{align}
while the spacelike sector is exactly isospectral. The qualitative distinction between the SR/spacelike case and the
timelike-type deformations is that the linear-in-$E$ term produces an additive displacement of both branches, thereby breaking
the antisymmetry $e_{n,-}=-e_{n,+}$.

\paragraph{Large-$n$ behaviour (fixed $\Omega$, small $\epsilon$).}
At fixed $\Omega$ and $\epsilon\ll 1$, the undeformed levels grow as $|e_{n,\pm}^{\rm SR}|\sim \sqrt{2n\Omega}$, whereas the
leading DSR shifts stay approximately constant, $\Delta e\sim -\epsilon/2$ in the first-power model and $\Delta e\sim -\epsilon$
in the MS model. The relative correction $|\Delta e|/|e_{n,+}^{\rm SR}|$ therefore decreases as $n^{-1/2}$ for sufficiently large
$n$, even though the absolute displacement is $n$-independent at leading order.

\subsection{Numerical illustration and figure discussion}
Because the realistic ratio $\epsilon=m/\Ep$ is tiny, we adopt an illustrative parameter set
\begin{equation}
\Omega=0.10,\qquad \epsilon=0.20,\qquad n=0,\ldots,25,
\label{eq:plotparams}
\end{equation}
which keeps all positive-branch levels comfortably below the timelike and MS pole region.

\begin{figure}[h!]
\centering
\includegraphics[width=0.92\linewidth]{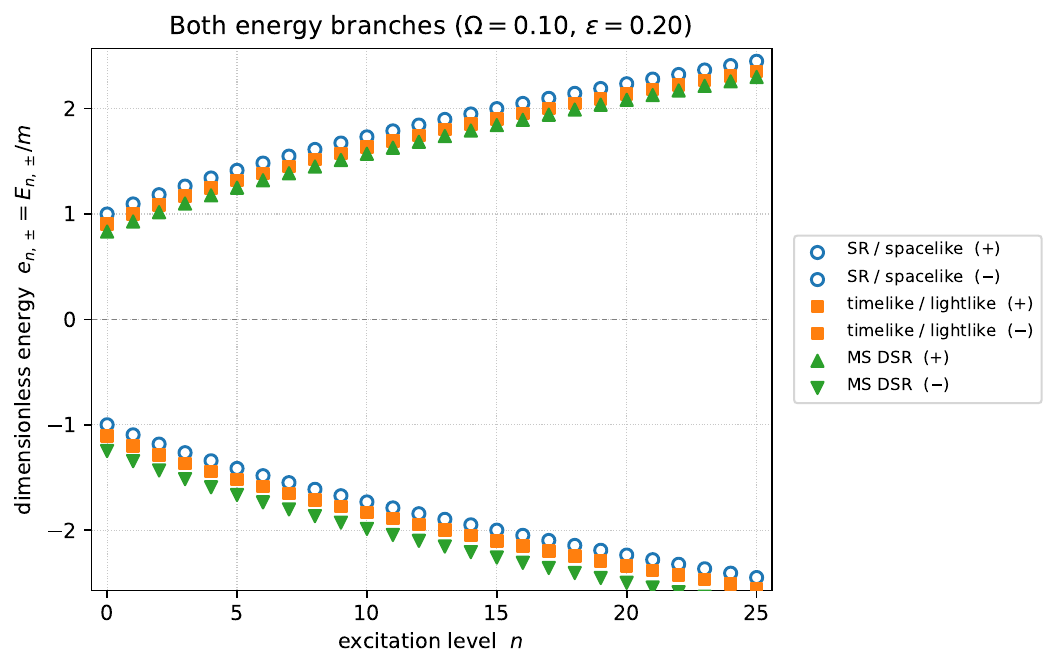}
\caption{\secondrev{Both energy branches $e_{n,\pm}=E_{n,\pm}/m$ as functions of the discrete excitation level $n$ for the
special-relativistic/spacelike spectrum, the timelike/lightlike first-power mass-shell invariant, and Magueijo--Smolin DSR,
using \eqref{eq:plotparams}. In the SR/spacelike case the spectrum is symmetric under $e\to-e$. In the timelike/lightlike and
MS cases, the linear-in-$E$ deformation shifts both branches by the same additive amount and breaks the exact relation
$e_{n,-}=-e_{n,+}$. The MS displacement is stronger at fixed $\epsilon$, consistently with its squared denominator.}}
\label{fig:bothbranches}
\end{figure}

\begin{figure}[h!]
\centering
\includegraphics[width=0.92\linewidth]{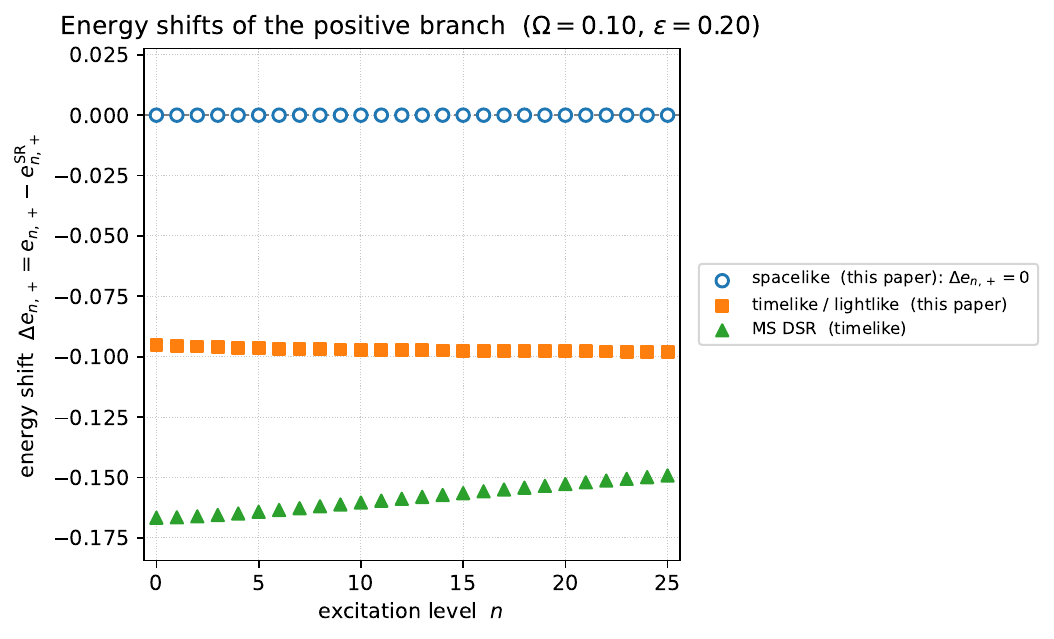}
\caption{\secondrev{Positive-branch energy shifts relative to special relativity,
$\Delta e_{n,+}=e_{n,+}-e_{n,+}^{\mathrm{SR}}$, for the parameters in \eqref{eq:plotparams}. The spacelike geometry is
exactly isospectral, $\Delta e_{n,+}=0$. The timelike and lightlike results coincide and show a negative Planck-suppressed
shift $\Delta e_{n,+}\approx-\epsilon/2$, whereas the Magueijo--Smolin model gives a larger negative shift,
$\Delta e_{n,+}\simeq-\epsilon$, at leading order.}}
\label{fig:shifts}
\end{figure}

\paragraph{Figure-level takeaway.}
Figure~\ref{fig:bothbranches} visually separates two distinct ``signatures'' of the deformation: (i) a pure displacement of both
branches (timelike, lightlike, and MS), and (ii) a pure wavefunction deformation at fixed spectrum (spacelike).
Figure~\ref{fig:shifts} then isolates the displacement itself and highlights the factor-of-two difference between the
first-power and the squared-denominator models at the same $\epsilon$. These figures make it explicit that, within solvable
bound-state systems, the power of the denominator is not a cosmetic modification but controls an in-principle spectral diagnostic
effect, especially within analogue realizations or effective models.

\section{Outlook}

The Dirac oscillator is a canonical relativistic oscillator model obtained from the Dirac equation by a nonminimal
substitution linear in position \cite{MoshinskySzczepaniak1989JPA,Quesne2017JPA}. It has deep connections with quantum optics
through exact mappings to Jaynes--Cummings-type models \cite{BermudezMartinDelgadoSolano2007PRA} and has even been realized
experimentally in analogue platforms \cite{FrancoVillafane2013PRL}. In $(1+1)$ dimensions, the squared Dirac oscillator
reduces to KG-type oscillator sectors with comparatively simple spin structure, suggesting that extensions of the present
analysis to spinorial DSR deformations are tractable.

\medskip
From a DSR perspective, two directions are particularly natural.
First, one can introduce the same linear--fractional mass-shell-invariant deformation at the Dirac level, then analyze how the resulting
linear-in-$E$ (or linear-in-$p$) terms reorganize the usual Dirac-oscillator ladder structure and spinor component coupling.
Because the KG oscillator already isolates the purely orbital sector, the present results provide a baseline for predicting
which deformations primarily cause \emph{energy-origin shifts} (timelike/lightlike) and which primarily cause \emph{non-Hermitian
representations} (spacelike-type).

Second, it would be valuable to combine the present MDR deformation with deformed phase-space structures motivated by
$\kappa$-Poincar\'e symmetry. In such a setting not only the dispersion relation but also the canonical commutators and the
operator realizations may be modified, so that the interplay with pseudo-Hermiticity becomes nontrivial. In particular, it would
be informative to determine when $\mathcal{PT}$ symmetry remains unbroken and to identify the corresponding metric operators
systematically for oscillator-type models.

\medskip
More broadly, the analytic solvability of the present $(1+1)$-dimensional system suggests several natural extensions:
(i) higher dimensions, with possible degeneracy lifting;
(ii) external fields (electric or magnetic) that compete with the oscillator scale;
and (iii) finite-temperature or thermal-field-theory versions, in which the branch asymmetry can affect thermodynamic quantities.
Each of these extensions may be sensitive to the denominator power in the invariant and could therefore provide additional
criteria for discriminating between DSR realizations in effective models.

\section{Conclusion}

We solved the KG oscillator for a family of linear--fractional (first-power) deformed mass-shell invariants and classified the
results by timelike, spacelike, and lightlike momentum-space geometries.

\medskip
\paragraph{Main spectral outcomes.}
The timelike and lightlike deformations yield identical closed-form spectra, characterized by a Planck-suppressed additive
shift that breaks the exact particle--antiparticle symmetry: both branches are displaced by $-m^2/(2\Ep)$ at leading order,
while the branch separation remains governed primarily by the usual oscillator scale $\sqrt{m^2+2nm\omega}$. The spacelike
deformation is exactly isospectral, so the entire discrete set of oscillator energies is unchanged with respect to the
special-relativistic case.

\medskip
\paragraph{Operator-theoretic and wavefunction outcomes.}
Although the spacelike spectrum remains real and undeformed, the associated spatial operator is non-Hermitian in the standard
$L^2(\mathbb{R})$ inner product, and the eigenfunctions acquire a complex translation together with a plane-wave phase factor.
We showed explicitly that this is a realization of an unbroken $\mathcal{PT}$-symmetric sector, and we provided a constructive
pseudo-Hermitian formulation: an explicit similarity map onto a Hermitian oscillator, the corresponding metric operator, and
the resulting $\eta$-orthogonality and biorthonormal structure. The lightlike realization combines the timelike-type spectral
displacement with the spacelike-type wavefunction shift, which makes it a convenient hybrid laboratory for separating
``energy-shift'' effects from ``state-deformation'' effects.

\medskip
\paragraph{Comparison with MS and significance.}
Comparison with the Magueijo--Smolin DSR model highlights the role of the denominator power in the deformed invariant: the MS
squared invariant produces a leading displacement twice as large as the first-power model at fixed $m/\Ep$, and also induces
additional higher-order distortions not reducible to a pure energy-origin shift. This shows that, in solvable bound-state
problems, the detailed functional form of the MDR (not merely the existence of a Planck scale) controls spectral
diagnostics in effective or analogue models.

\medskip
\paragraph{\secondrev{Scope and limitations.}}
\secondrev{The present analysis is subject to several limitations that define the domain of validity of the conclusions. First, the
investigation is restricted to a stationary Klein--Gordon oscillator in $(1+1)$-dimensional spacetime. Although this setting
permits an exact analytical treatment and provides a controlled framework for comparing different deformations of the
relativistic mass-shell condition, extensions to higher dimensions may introduce additional degeneracy patterns, angular
structures, and spin-dependent effects. Second, the DSR deformation is implemented through a nonlinear momentum-space map
and the associated modified mass-shell invariant. A complete formulation incorporating the corresponding Hopf-algebra
structure, a deformed momentum-composition law, or consistently modified phase-space commutators lies beyond the scope of
the present work. The analysis is also restricted to the physical branch $1+\lp\,a\cdot p>0$, on which the nonlinear map is
real, regular, and continuously connected to the special-relativistic limit. Finally, the comparatively large deformation
ratio used in the numerical illustrations is chosen solely to make the qualitative spectral differences visible. For
realistic particle masses and a Planck-scale deformation energy, the predicted corrections are strongly suppressed.
Consequently, the spectra obtained here should primarily be regarded as exact theoretical benchmarks for identifying the
structural consequences of the considered DSR realizations, rather than as direct phenomenological predictions within
current experimental sensitivity.}

\medskip
\paragraph{Outlook.}
The present analysis provides a controlled benchmark for studying DSR-inspired deformations in interacting relativistic
systems. Extensions to Dirac oscillators, higher dimensions, and external fields appear particularly promising, especially
for clarifying when pseudo-Hermiticity and $\mathcal{PT}$ symmetry protect spectral reality and for determining how different
nonlinear realizations of DSR reorganize operator algebras beyond the free-particle sector.

\appendix

\section{Metric operator and $\eta$-orthogonality in the spacelike/lightlike sectors}
\label{app:metric}

Let $h$ denote the Hermitian oscillator operator obtained after the complex shift and momentum shift, and let
$H$ denote the original non-Hermitian operator in \eqref{eq:SLexpanded}. The similarity map in the main text reads
\begin{equation}
H=\mathcal{S}^{-1}h\,\mathcal{S},
\qquad
\mathcal{S}=e^{-\delta \hat{p}}e^{-\ii\kappa x},
\end{equation}
with $\delta,\kappa$ given in \eqref{eq:kappadelta}. Since $h^\dagger=h$, we compute
\begin{equation}
H^\dagger
=
\left(\mathcal{S}^{-1}\right)^\dagger h \,\mathcal{S}^\dagger
=
\left(\mathcal{S}^\dagger\right)^{-1}h\,\mathcal{S}^\dagger
=
\left(\mathcal{S}^\dagger\mathcal{S}\right)\,H\,\left(\mathcal{S}^\dagger\mathcal{S}\right)^{-1}.
\end{equation}
Therefore $H$ is \emph{pseudo-Hermitian} with respect to the positive operator
\begin{equation}
\boxed{
\eta:=\mathcal{S}^\dagger\mathcal{S}
}
\label{eq:etaDef}
\end{equation}
in the sense that
\begin{equation}
\boxed{
H^\dagger=\eta\,H\,\eta^{-1}.
}
\end{equation}

\paragraph{Physical inner product and orthogonality.}
Define the $\eta$-inner product by
\begin{equation}
(\psi,\varphi)_\eta := \langle \psi|\eta|\varphi\rangle.
\end{equation}
If $\{|n\rangle\}$ are the orthonormal eigenstates of $h$ with $\langle m|n\rangle=\delta_{mn}$ and we define
\begin{equation}
|\psi_n\rangle := \mathcal{S}^{-1}|n\rangle,
\end{equation}
then
\begin{equation}
(\psi_m,\psi_n)_\eta
=
\langle \psi_m|\eta|\psi_n\rangle
=
\langle m|\left(\mathcal{S}^{-1}\right)^\dagger \mathcal{S}^\dagger\mathcal{S}\,\mathcal{S}^{-1}|n\rangle
=
\langle m|n\rangle
=
\boxed{\delta_{mn}}.
\end{equation}
Thus the shifted states are orthonormal in the $\eta$-metric even though they are not orthonormal in the standard
$L^2(\mathbb{R})$ product. A convenient associated dual set is
\begin{equation}
|\chi_n\rangle := \eta|\psi_n\rangle,
\end{equation}
so that the corresponding left eigenbras are $\langle\chi_n|=\langle\psi_n|\eta$ and satisfy the biorthonormal
relations $\langle \chi_m|\psi_n\rangle=\delta_{mn}$.
These constructions are standard in the pseudo-Hermitian/$\mathcal{PT}$-symmetric framework
\cite{BenderBoettcher1998PRL,Mostafazadeh2002JMP,Mostafazadeh2003JMP}.
\section*{Funding information}

Funding: Not applicable.
 
\bigskip
\bibliographystyle{unsrt}
\bibliography{references}

\end{document}